# Ordering intermetallic alloys by ion irradiation:
# a way to tailor magnetic media

**H. Bernas[1*], D. Halley[2], K.-H. Heinig[3], J.-Ph. Attané[2],**

**D. Ravelosona[4], A. Marty[2], P. Auric[2], C. Chappert[4], Y. Samson[2]**

[1] *Centre de Spectrométrie Nucléaire et de Spectrométrie de Masse,*

*CNRS-Université Paris XI, 91405 – Orsay, France,*

[2] *CEA Grenoble, DRFMC – Service de Physique des Matériaux & Microstructures,*

*(assoc. with Université J. Fourier) 38054 – Grenoble, France,*

[3] *FZ-Rossendorf, Inst. Ionenstrahlphys. & Materialforsch., 01314 – Dresden, Germany,*

[4] *Inst. d'Electronique Fdtale., CNRS–Université Paris XI, 91405 – Orsay, France*

Combining He ion irradiation and thermal mobility below 600K, we both trigger and control the transformation from chemical disorder to order in thin films of an intermetallic ferromagnet (FePd). Kinetic Monte Carlo simulations show how the initial directional short range order determines order propagation. Magnetic ordering perpendicular to the film plane was achieved, promoting the initially weak magnetic anisotropy to the highest values known for FePd films. This post-growth treatment should find applications in ultrahigh density magnetic recording.

*\* Corresponding author (email: bernas@csnsm.in2p3.fr)*

Atomic collisions in solids are often associated with the concept of disorder. In fact, the mobility induced in alloys by ion irradiation at appropriate temperatures can produce a wealth of phases[1] which may (or not) be related to the equilibrium phase diagram. Research on such "driven alloys"[2] is quite active. Here, we demonstrate that irradiation may even transform chemical disorder into order in intermetallic alloys. Our chosen example is the ferromagnetic $L1_0$-structure alloy FePd, where chemical order corresponds to a layered structure with alternating Fe and Pd planes along the c-axis of a tetragonal structure. Their very large magnetic anisotropy energy makes FePd (or isostructural FePt) films excellent candidates for future magnetic media[3], if only uniaxial ordering with the easy magnetic axis parallel (or perpendicular) to the surface may be achieved. Whereas codeposited films order at growth temperatures ( 700K for FePt[4], ~650K for FePd[5]) which are prohibitive[6] for media processing, we show that an ion irradiation process[7] operating at ~550K on chemically disordered FePd films allows a controlled transition to $L1_0$ order. The technique should also apply to other layered-structure intermetallics.

The possibility of irradiation-induced ordering was conjectured with impressive insight by Néel[8], and then forgotten. Recently[9], significant structural and magnetic order was obtained and controlled in FePt by post-growth He ion irradiation at 570 K, well below the ordering temperature, but the initial sample preparation's influence on final long range order (LRO) was not understood[10]. Focussing on FePd films grown by a "layer-by-layer" technique[11] which produces an anisotropy bias in the film plane, we show that practically complete ordering perpendicular to the film plane may be obtained by irradiation, if directional short range order (DSRO) is initially present.

Three requirements must be satisfied by the irradiating beam: small energy transfers, minimizing recoil displacements; low collision cross-section in order to avoid defect interactions; sufficient beam energy to insure that ions stop well inside the substrate. Using He ions at energies around 30-130 keV[12], recoils are limited[13] to one (or rarely two) atomic distance(s), the number of displaced atoms in the alloy is ~ $5.10^{-2}$ / incoming ion / nm, and all He ions stop in the substrate below the deposited layers.

About one Frenkel (associated interstitial and vacancy) pair is thus introduced in the entire film per incident ion. Over half of these pairs recombine athermally[14]; a small fraction may lead to antisite defect formation in the ordered phase. Pairs also recombine thermally, but due to our thin film geometry most (very mobile) interstitials disappear at surfaces. We are left with a fraction of surviving vacancies reaching some 10% (interstitials will henceforth be neglected). They induce FePd ordering by successive individual atomic rearrangements that are determined by the Fe-Fe, Pd-Pd and Fe-Pd binding energies. The system thus explores the nonequilibrium paths towards lower energy configurations corresponding to chemical order, with a vacancy jump probability that depends exponentially on the ratio of the binding energies to kT.

The "layer-by-layer" FePd films were obtained[11] by masking the Fe or Pd atomic fluxes in a molecular beam epitaxy system, growing a biatomic period multilayer (typically 10-50 nm) of alternating pure Fe and pure Pd atomic planes on a MgO(001) substrate at 300 K covered by a 3 nm Cr seed layer and a 60 nm (001) Pd buffer layer. Oxidation was avoided by a 2 nm Pd capping layer.

X-ray diffraction intensities of the (001) and (003) superstructure peaks identify the variant of the $L1_0$ ordered structure with the quadratic c-axis perpendicular to the film surface. Integrating these peaks and the fundamental (002) and (004) peaks, we calculated[11] the LRO parameter S for the films (defined as S= $n_{Fe}$ - $n_{Pd}$, where n is the site occupancy on the corresponding sublattices). The initial LRO was weak, typically S=0.1. Polarized EXAFS measurements[11] revealed that this did not preclude a high degree of DSRO and significant in-plane homo-coordination. The films were irradiated at temperatures ranging from 293K to 573K with 130 keV He ions at a fluence of $2.0x10^{16}$ ions/cm$^2$. Partially masked samples confirmed that heating alone had no effect. **Fig. 1, upper part** shows $L1_0$ long-range ordering (from S ~ 0.1 to ~ 0.65), obtained after irradiation at 573K.

Irradiation at 573 K led to a rotation of the easy magnetization axis from in plane to out of plane (**Fig.1, lower part**). The quality factor Q=$K_u$/[(1/2)$\mu_0 M_s^2$] that compares the uniaxial anisotropy $K_u$ to the shape anisotropy (where $M_s$= 1,05.10$^6$ A.m$^{-1}$) may be derived from in-plane and out-of-plane hysteresis loops. It increased from 0.76 (easy axis in-plane) to 1.5 ± 0.1 (easy axis out-of-plane) after irradiation at 473 K or 523 K; the perpendicular saturation and nucleation fields are both correspondingly reduced. The same experiments on layer-by-layer grown isostructural FePt layers showed similar changes, with the perpendicular magnetic remanence even increasing from below 5% to 100%.

Direct information on spin ordering was obtained (**Fig. 2**) from Mössbauer spectra (MS) before and after irradiation. In our geometry ( -ray detection axis perpendicular to the film), we detect the perpendicular spin alignment via the line intensities at ± 3.6 mm/s. Whereas the initial chemically disordered sample (A) had a near-zero $L1_0$ component perpendicular to the film plane, the latter (B) reached 60-70% after irradiation. The MS is then very similar to that (C) of a high-temperature deposited sample[15] with the highest known perpendicular anisotropy for FePd. MS of FePt films show the same, most striking, confirmation of irradiation-induced ordering.

Kinetic lattice Monte Carlo (KLMC) simulations confirmed the influence of pairwise interactions and of our initial structural conditions on ordering. They were performed in a 256x64x64 atom box, comparable to the actual film thickness, on an fcc lattice with an equiatomic average composition. In order to simplify KLMC calculations, the true $L1_0$ tetragonal structure was not introduced. Atomic size effects are thus neglected, but since the corresponding energies are much smaller than the ordering energy, this does not affect our conclusions regarding ordering. As mentioned, the kinetics are entirely due to vacancy motion. The film being thin, Frenkel pair production being very low and the ion flux weak, vacancies act one by one – their activation energy and diffusion constant play no role: the final degree of order and the ordering rate depend only on the vacancy ordering efficiency. Energy barriers between Fe and Pd atoms were deduced[16] from the elements' cohesive energies: the bond strength values used here were 0.70 eV (Fe-Fe), 0.65 eV (Pd-Pd), and 0.78 eV (Fe-Pd). In the absence of a measured value for FePd, we used the fcc(A1)   $L1_0$ ordering enthalpy of FePt[17] as an upper limit. Boundary conditions were periodic in the horizontal plane. Layers of pure Pd were introduced above and below the FePd layer to account for the buffer and capping layers, and the upper surface and the bottom Pd/MgO interface (which contains dislocations[5]) were sinks for diffusing vacancies. The KLMC code was described elsewhere[18]. The initial repartition of Fe and Pd atoms was taken from the DSRO values of Gehanno et al[11]: starting from a central Fe atom, the probabilities of finding a Fe (Pd) atom in the plane are respectively 0.82 (0.18), and out of the plane they are resp. 0.39 (0.61). **Fig. 3** shows snapshots of the ordering process. We note that (1) The driving force for ordering is the energy gained by maximizing the number of Fe-Pd bonds, and this nearest-neighbor scale mechanism is clearly effective. Directional ordering then depends on the existence and size of a pure Fe (or Pd) layer "template". The DSRO anisotropy plays a crucial role in promoting ordering along the variant perpendicular to the film plane. Initial film growth conditions are critical insofar as they determine template features. (2) Layer defects (e.g., bilayer islands of Fe or Pd) are quickly erased under vacancy motion for the reason discussed above, and site-antisite pairs are eliminated even more efficiently. (3) $\{100\}$ antiphase boundaries develop as the $[001]$ - oriented domains grow. Their low energy (in the meV range) allows them to survive. (4) $[001]$ growth dominates by far in Fig. 3 because of the initial DSRO anisotropy, but we checked that if the latter is reduced - as in samples prepared via codeposition or sputtering[9,10]– domains with other variants may interfere and the Q-factor would be correspondingly reduced. This will be detailed elsewhere.

In summary, because He ion irradiation-induced rearrangements are on a near-neighbor scale, successive pairwise interactions lead to efficient ordering. Simulations show that ordering proceeds from the initial DSRO, via the energy gained by maximizing the number of Fe-Pd bonds. The final degree of order and the ordering rate depend on the initial DSRO. Using this information and taking advantage of the "local" nature (spin-orbit dependence) of the magnetic anisotropy, we obtain films with perpendicular anisotropy values as large as those measured in bulk FePd. Our process therefore provides a post-growth method to orient high density magnetic media, and should also apply to nanoparticulate media whose narrow size dispersion is otherwise difficult to preserve[19] during their high-temperature L10 ordering process. More generally, by playing with competing bonding strengths and starting from deposited films with appropriate DSRO, ion beam ordering of intermetallics may also be envisaged as a new tool in thin film growth technology and surface property improvement.

We thank O. Kaitasov and S. Gautrot for their technical contributions, K. Barmak for a discussion and R. Maynard for drawing our attention to Ref. 9. This work was partially supported by the ISARD program (Université Paris XI), and the French Ministry of Research under ACI "Nanorad". HB is grateful to IfIM/FZ – Rossendorf for hospitality and support.

## FIGURE CAPTIONS

**Fig. 1 – (upper part)** Irradiation temperature dependence of (001) diffraction peak intensity for a 40 nm thick layer-by-layer FePd film deposited at room temperature. Peak intensities are normalized to the (002) peak. Irradiation by 130 keV He ions at a fluence of $2.10^{16}$ ions.cm$^{-2}$. [From Ref. 10]

**(lower part)** : The in-plane (//) and out-of-plane ( ) hysteresis loops for the same sample. Before irradiation (a) the easy magnetization axis is in-plane, whereas after irradiation at 623 K (b) it has rotated from in-plane to out-of-plane.

**Fig. 2** : Mössbauer spectra of a FePd sample similar to that of Fig. 1, taken before **(A)** and after **(B)** irradiation at 623 K by 130 keV He ions (fluence $2.10^{16}$ ions.cm$^{-2}$). Spectrum B is almost identical to the spectrum of a high temperature (700K) codeposited sample **(C)**, whose perpendicular anisotropy (Q= 1.8, close to the highest value for bulk FePd) corresponds to nearly complete chemical ordering (S=0.8). Lines are fits to hyperfine field distributions (see Ref. 15).

**Fig. 3** : Snapshots from KLMC simulation showing vacancy-induced "L1$_0$" chemical ordering during He ion irradiation at 550 K of a 60 nm FePd film sandwiched between a Pd buffer layer and a Pd capping layer. The figures are cuts (2 atomic planes thick) through the simulation cell - for clarity, only Pd atoms are shown (in black). The (001) orientation of the L1$_0$ structure is in the z direction. The initial structure (first snapshot on left) incorporates the experimentally determined anisotropic DSRO from Ref. 12. While ordering can occur along the three possible variants (horizontal and vertical stripes correspond respectively to the variants along the z and x axes, and checkerboards to the variant in the y direction), due to the DSRO the z-axis variant (i.e.: "L1$_0$" order

along the direction perpendicular to the film) rapidly dominates. Note the presence of antiphase boundaries perpendicular to the film. The total number of vacancies introduced into the cell is shown on each snapshot. Complete ordering (third snapshot) corresponds to a vacancy input of about $2 \times 10^{15}$ per $cm^2$, in agreement with the experimental value.

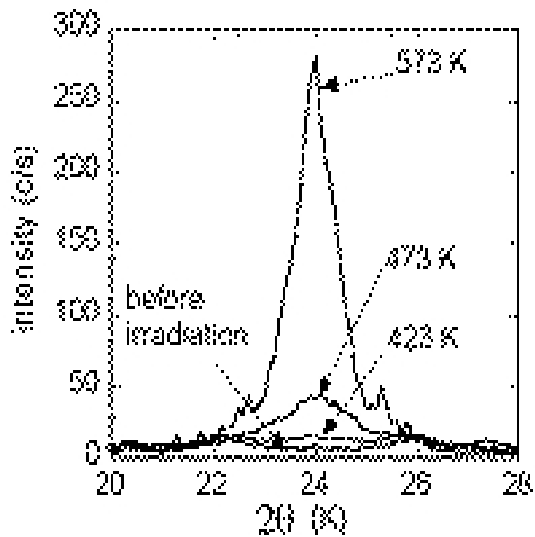

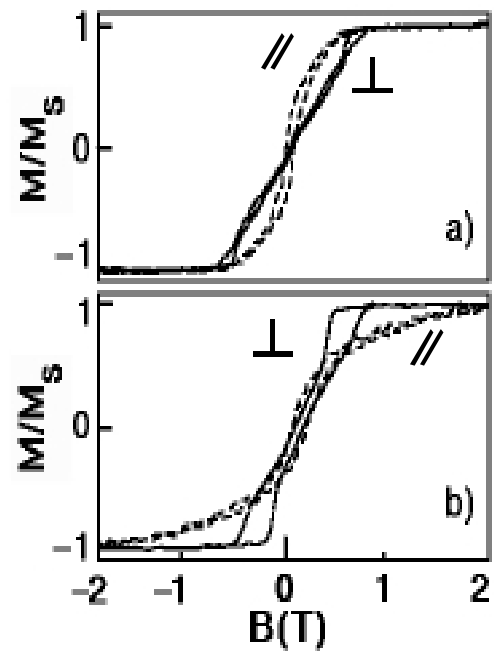

Fig1- upper part                          Fig. 1 (lower part)

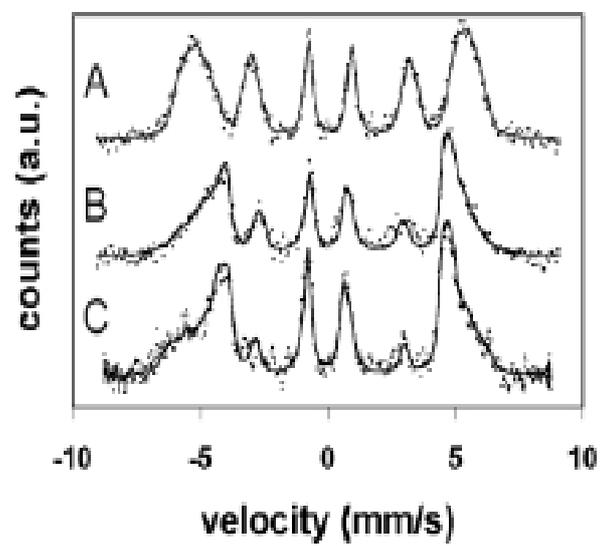

Fig. 2

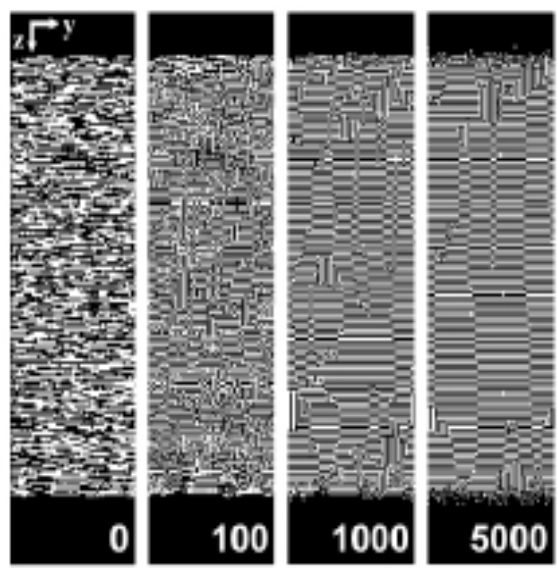

Fig. 3